\documentclass[aps,prl,twocolumn,showpacs,amsmath,amssymb,floatfix,
superscriptaddress]{revtex4}
\usepackage{graphicx}

\newcommand {\be}{\begin{equation}}
\newcommand {\ee}{\end{equation}}

\begin{document}

\title{Collective dynamics in sparse networks}
\date{\today}

\author{Stefano Luccioli}
\affiliation{CNR - Consiglio Nazionale delle Ricerche - Istituto dei Sistemi Complessi, via Madonna del Piano 10, I-50019 Sesto Fiorentino, Italy}
\affiliation{INFN Sez. Firenze, via Sansone, 1 - I-50019 Sesto Fiorentino, Italy}
\affiliation{Centro Interdipartimentale per lo Studio delle Dinamiche Complesse, via Sansone,   1 - I-50019 Sesto Fiorentino, Italy}

\author{Simona Olmi}
\affiliation{CNR - Consiglio Nazionale delle Ricerche - Istituto dei Sistemi Complessi, via Madonna del Piano 10, I-50019 Sesto Fiorentino, Italy}
\affiliation{INFN Sez. Firenze, via Sansone, 1 - I-50019 Sesto Fiorentino, Italy}
\affiliation{Centro Interdipartimentale per lo Studio delle Dinamiche Complesse, via Sansone,   1 - I-50019 Sesto Fiorentino, Italy}

\author{Antonio Politi}
\affiliation{Institute for Complex Systems and Mathematical Biology and SUPA, University of Aberdeen, Aberdeen AB24 3UE, United Kingdom}
\affiliation{CNR - Consiglio Nazionale delle Ricerche - Istituto dei Sistemi Complessi, via Madonna del Piano 10, I-50019 Sesto Fiorentino, Italy}
\affiliation{Centro Interdipartimentale per lo Studio delle Dinamiche Complesse, via Sansone,   1 - I-50019 Sesto Fiorentino, Italy}

\author{Alessandro Torcini}
\affiliation{CNR - Consiglio Nazionale delle Ricerche - Istituto dei Sistemi Complessi, via Madonna del Piano 10, I-50019 Sesto Fiorentino, Italy}
\affiliation{INFN Sez. Firenze, via Sansone, 1 - I-50019 Sesto Fiorentino, Italy}
\affiliation{Centro Interdipartimentale per lo Studio delle Dinamiche Complesse, via Sansone,   1 - I-50019 Sesto Fiorentino, Italy}

\begin{abstract}
The microscopic and macroscopic dynamics of random networks is investigated
in the strong-dilution limit (i.e. for sparse networks). By simulating
chaotic maps, Stuart-Landau oscillators, and leaky integrate-and-fire neurons,
we show that a finite connectivity (of the order of a few tens) is able to
sustain a nontrivial collective dynamics even in the thermodynamic limit.
Although the network structure implies a non-additive dynamics,
the microscopic evolution is extensive (i.e. the number of active degrees of
freedom is proportional to the number of network elements).
\end{abstract}


\maketitle

The organization of dynamical phenomena on different scales is 
a general property of systems out-of-equilibrium, such as those
encountered in plasma physics, turbulence, and neuroscience. The simplest
instance of this hierarchical organization is the spontaneous emergence of
collective behaviour out of a microscopically chaotic dynamics, a phenomenon
reminiscent of equilibrium phase-transitions.
The first studies of collective dynamics contributed to uncover time-dependent 
macroscopic states with different degrees of complexity in mean field models
\cite{kaneko0,kaneko} and in spatio-temporal chaotic models \cite{manneville}.

Complex networks provide an even more interesting setup for the study of 
macroscopic phases, since this is the typical structure of many non-equilibrium
systems. Most of the studies of network dynamics have been so far devoted to
the characterization of synchronized
regimes~\cite{review_syncnet,grinstein}, where the single oscillators evolve in a
coherent way. However, some preliminary studies, especially of neural
networks with stochastic noise~\cite{brunel_hakim1999}, have shown that
self-sustained macroscopic oscillations can spontaneously arise also when
the single elements evolve in a seemingly uncorrelated way.
Altogether, the emergence of collective dynamics has been investigated in
the presence of various ingredients such as delayed interactions, diversity of 
the single elements, time-dependent synaptic
connections~\cite{timme_synch_delay,denkel_synch_heterogeneity,maistrenko_kuramoto_STDP}.
In particular, it is known that disorder may give rise to an extremely rich
macroscopic scenario: this is indeed the framework where glassy phenomena have 
been uncovered \cite{glass} and a highly irregular dynamics
observed in neural networks \cite{nostro_1}.

In this Letter, we study several random networks to clarify the role played
by the (in-degree) connectivity $K$ (i.e. the number of incoming connections per node)
on the onset of collective motion. It is convenient to distinguish between two
classes of systems~\cite{GolombHanselMato_2001}: {\it massive} networks, where $K$ is 
proportional to the network size $N$; {\it sparse} (or strongly diluted) networks, 
where $K \ll N$, and specifically $K$ is independent of $N$ as $N \to \infty$. 
Lattice systems with short-range interactions belong to the
latter class. While it is not surprising to observe the onset of a collective motion
in massive networks, it is less obvious to predict whether and when this can happen
in sparse ones. In a model of leaky integrate-and-fire (LIF) neurons,
it has been shown that a finite connectivity can sustain a partially synchronized
regime \cite{hansel}. Here, we show that the emergence of a collective
dynamics above a finite critical connectivity $K_c$ is a general and robust property
of sparse networks of oscillators. Since $K_c$ turns out to be of the order of a few
tens in all models we have investigated, macroscopic motion appears to be rather 
ubiquitous and possibly relevant in the context of neural dynamics. In our simulations,
we have typically assumed that all nodes are characterized by the same connectivity $K$, 
but we have verified that the same scenario holds assuming  a Poissonian degree
distribution with average connectivity $K$, as in Erd\"os-Renyi graphs.

Finally, we analyse the microscopic dynamics, irrespective of the presence
of the macroscopic phase, finding that it is always extensive (the number of
unstable directions, as well as the power contained in the principal components,
is proportional to the network size).
This property is highly nontrivial, as the network dynamics is non additive
(it cannot be approximated with the juxtaposition of almost indepedent
sub-structures, see below). This is at variance with globally coupled systems,
which exhibit a non-extensive
component in the Lyapunov spectrum \cite{KazumasaPRL2011}.

More specifically we study three classes of dynamical systems: (i) units that
are chaotic by themselves (logistic maps - LM); (ii) units that may become
chaotic as a result of a periodic forcing (Stuart-Landau oscillators - SL);
(iii) phase-oscillators that cannot behave chaotically under any forcing
(LIF neurons).

{\it Coupled maps.}
The dynamics on a network of $N$ coupled logistic maps (LM) is defined as
\be
x_{n+1}(i)=(1-g)f(x_{n}(i))+ g h_n(i)  \qquad ,
\label{eq:cm}
\ee
where $x_{n}(i)$ represents the state of the {\it i}th node ($i=1,\dots,N$) at
time $n$, the logistic map $f(x)= ax(1-x)$ rules the internal dynamics, 
and $g$ is the coupling strength. Finally, 
$h_{n}(i) = (1/K)\sum_{j=1}^{N} S_{ij}f(x_{n}(j))$
denotes the local field, where $S_{ij}$ is the connectivity matrix:
$S_{ij}=1$ if an incoming link from $j$ to $i$ is present, otherwise
$S_{ij}=0$. It is convenient to introduce the average field
${\overline h}_n$ and its standard deviation $\sigma_h$
($\sigma_{h}^{2}=\langle {\overline h}_n^{2} \rangle - \langle {\overline h}_n
 \rangle^{2}$)~\cite{note} .

\begin{figure}[h]
\begin{center}
\includegraphics*[angle=0,width=7.5cm]{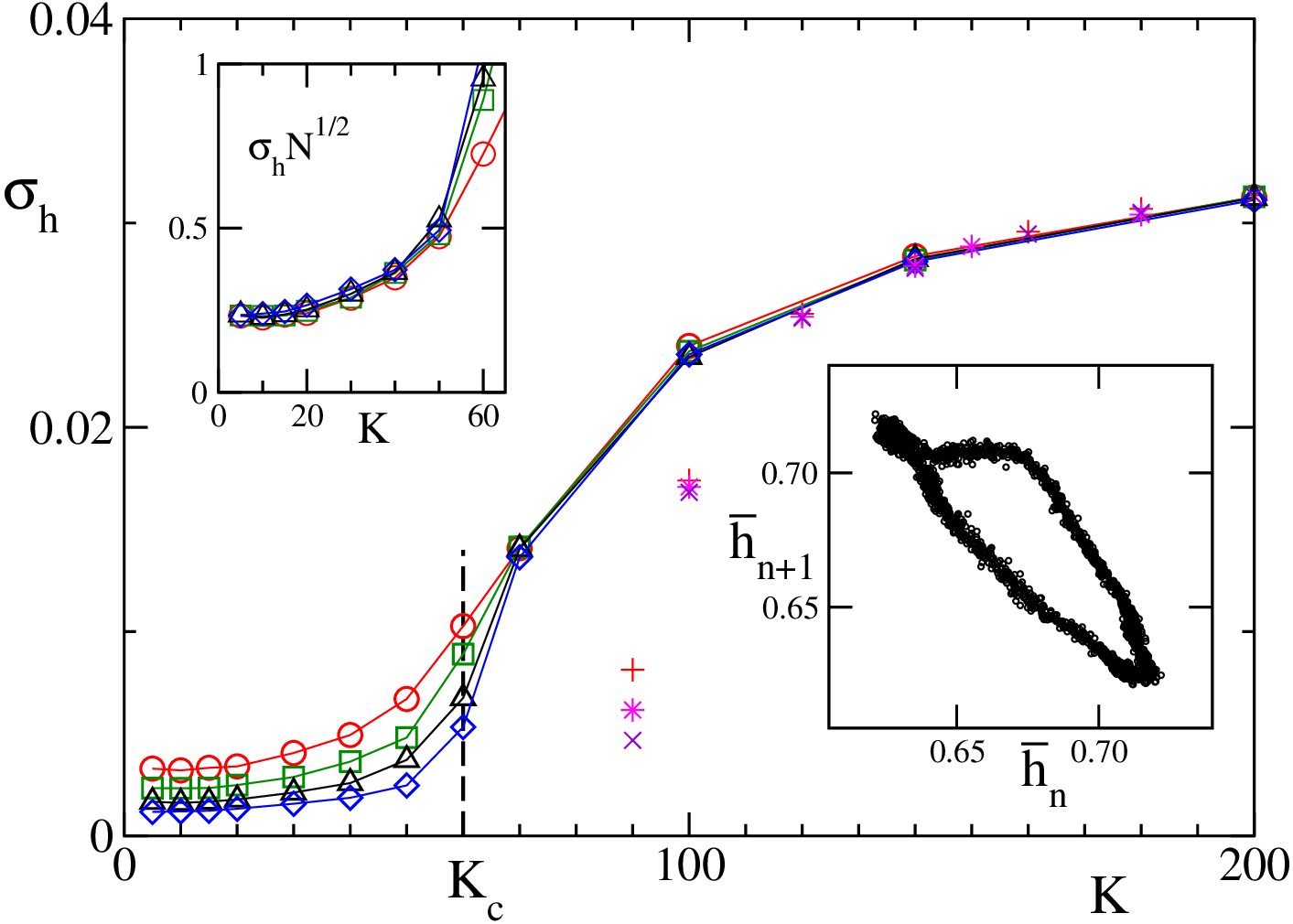}
\end{center}
\caption{(color online) Model LM. Standard deviation 
of the mean field, $\sigma_h$ (averaged over 5 realizations of disorder) 
versus $K$ for $N=5,000$
(red) circles, $N=10,000$ (green) squares, $N=20,000$ (black) triangles, 
and $N=40,000$ (blue) diamonds. The upper inset shows the rescaled 
$\sigma_h$ at low $K$. In the lower inset the return map of 
${\overline h}_n$ for $K=500$ 
and $N=20,000$. Symbols not connected by lines refer 
to $K=2 N_s+k_R$ ($N_s=40$ is the number of nearest neighbours
and $k_R$ of the random links)
for $N=5,000$ (violet) crosses, $N=10,000$ (magenta) stars and $N=20,000$ (red) plus.} 
\label{mappe:1}
\end{figure}

In Fig.~\ref{mappe:1}, $\sigma_h$ is plotted versus the connectivity $K$ for
$a=3.9$, $g=0.1$ and increasing network sizes. For low connectivity, $\sigma_h$
is quite small
and decreases as $1/\sqrt{N}$ with the system size (see left inset),
i.e. the deviation from zero is a finite-size effect. Above $K_{c}\simeq 60$,
$\sigma_h$ assumes finite values, independently of the system size,
signaling the onset of a collective dynamics. In fact, the right inset in
Fig.~\ref{mappe:1} reveals nontrivial collective oscillations for $K=500$ and
$N=20,000$ (we have verified that the thickness of the ``curve" does not
decrease upon increasing the network size - data not shown). The
phase-portrait is analogous to that previously obtained in globally coupled
maps \cite{kaneko}. This indicates that the evolution of a sparse
network reduces, for $K\to\infty$, to that of its corresponding mean-field
version. What is new and a priori non obvious is that a finite and relatively
small connectivity suffices to sustain a macroscopic motion.

\begin{figure}[h]
\begin{center}
\includegraphics*[angle=0,width=8cm]{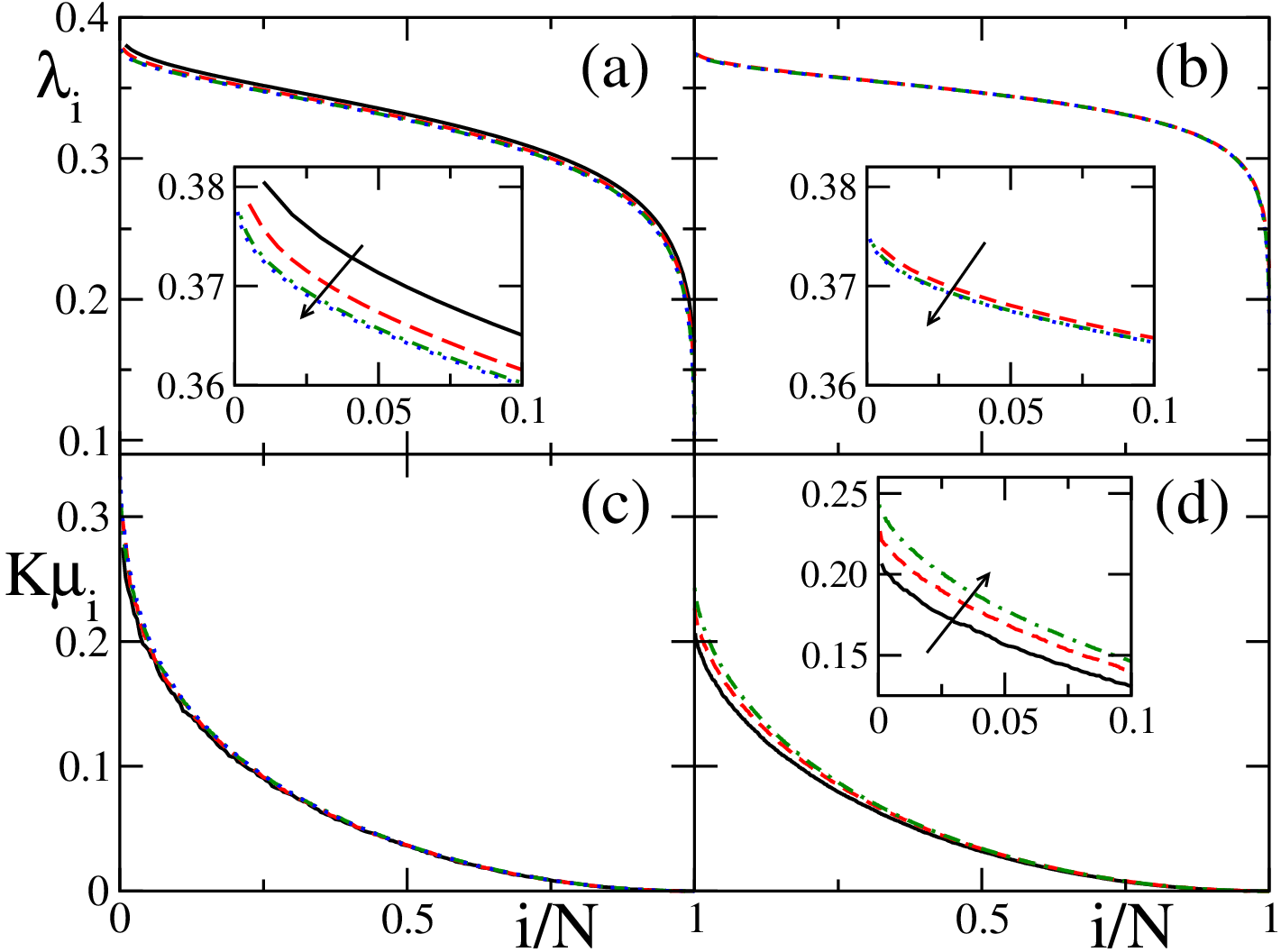}
\end{center}
\caption{(color online) Model LM. 
LE spectra are reported in (a) for $K=10$ 
and $N=100-200-500-1,000$, 
and in (b) for $K=80$ and $N=200-500-1000$. 
CE spectra (rescaled by $K$) are reported in 
(c) for $K=10$ and $N=200-400-800-1,600$, and in (d) 
for $K=100$ 
and $N=800-1,600-3,200$.
In the insets a zoom of the largest values is shown.
In this and in the following figures
the arrow direction indicates data obtained for increasing system sizes.
} 
\label{mappe:2}
\end{figure}

As for the evolution of the single units, the most appropriate tool to
investigate the microscopic dynamics is Lyapunov analysis. In
Fig.~\ref{mappe:2}a,b we can see that both below and above $K_{c}$ the
dynamics is characterized by extensive high-dimensional
chaos~\cite{grass89,extensivity}, since the spectra of the Lyapunov exponents
(LE) $\{\lambda_i\}$ collapse onto one another, when they are plotted versus
the intensive variable $i/N$ \cite{livi_ruffo_politi}. In the
inset, one can appreciate that the convergence occurs also for the largest LE,
at variance with the non-extensive behavior, recently detected in globally
coupled networks \cite{KazumasaPRL2011}.

In a sparse network, the field $h_n(i)$ fluctuates with $i$, no matter how large 
the network is, since $h_n(i)$ is the sum of a finite number of contributions.
One way to characterize its variability is by determining the covariance exponents
(CE), i.e. the eigenvalues $\mu_{i}$ of the matrix
$C_{ij}=\langle\delta h_n(i)\delta h_n(j)\rangle-\langle\delta h_n(i)\rangle
\langle \delta h_n(j) \rangle$, where 
$\delta h_n(i)$ = $h_n(i)- {\overline h}_n$.
In 1D spatial systems with periodic boundary conditions such an approach
would correspond to determine the spatial Fourier spectrum.
In this case, since there is no ``wavelength'' to refer to, it is natural
to order the eigenvalues from the largest to the smallest one. The results for different
network sizes are plotted in Fig.~\ref{mappe:2} versus $i/N$, (panel $c$ and $d$
refer to $K$ values below and above $K_c$, respectively). The good data collapse
confirms the extensivity of microscopic fluctuations. In both panels, $\mu_i$
has been rescaled by $K$, to emphasize extensivity in yet a different way; in
fact, as the local field is the sum of $K$ contributions, its variance is
expected to be on the order of $1/K$.

\begin{figure}[h]
\begin{center}
\includegraphics*[angle=0,width=8cm]{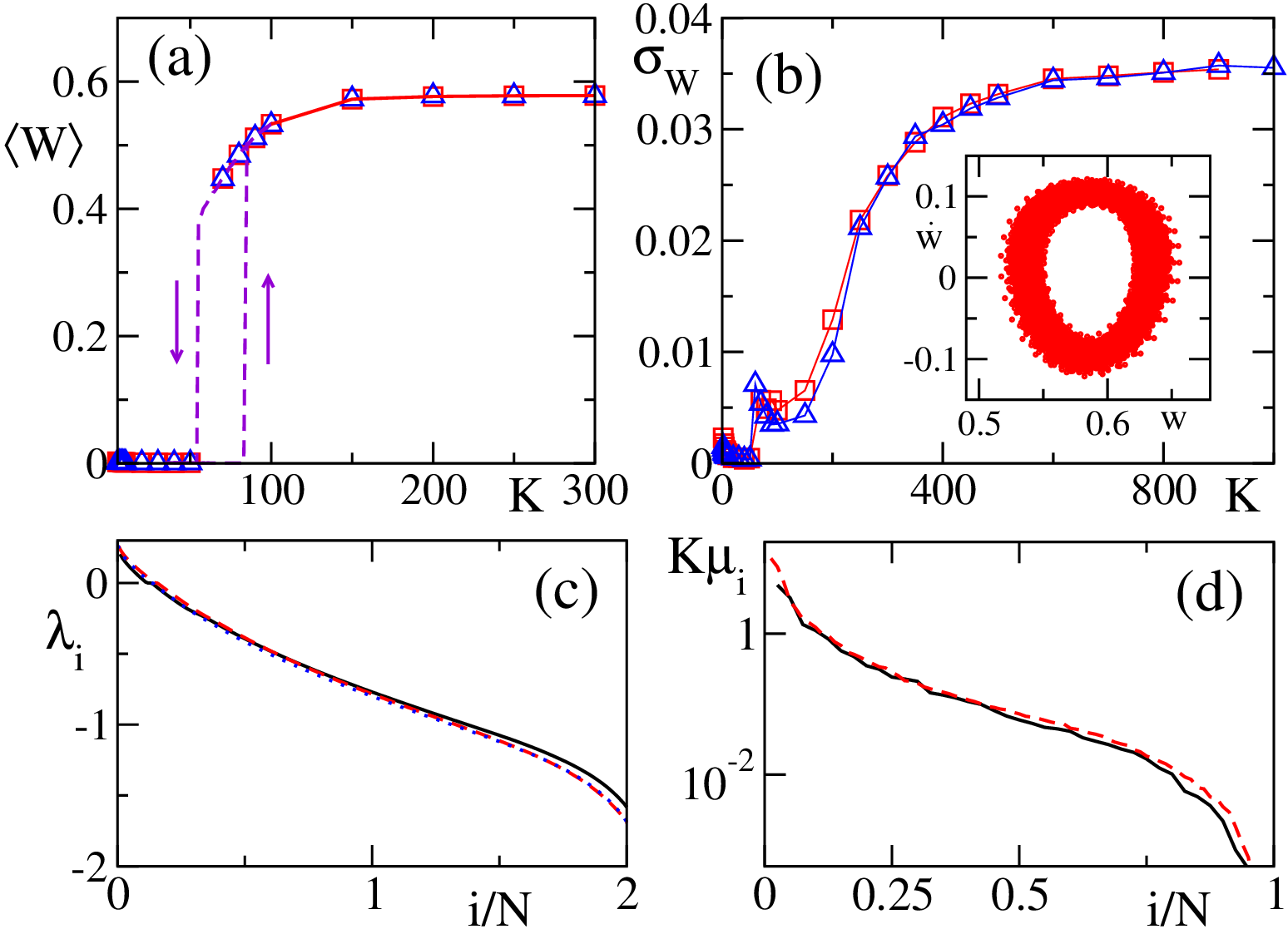}
\end{center}
\caption{(color online) Model SL. Average (a) and 
standard deviation (b) of $W$ 
versus $K$ for $N=10,000$ (red) squares, and $N=20,000$ (blue) triangles 
(averaged over 5 realizations of the disorder). 
The dashed lines in (a) show the region of bistability (see the text). The
inset in (b) shows the macroscopic attractor for $N=10,000$ and $K=800$.  
(c) LE spectra for $K=10$ and $N=100-200-400$ .
(d) CE spectra (rescaled by $K$) for $K=10$ and $N=40-80$. 
} 
\label{sl:1}
\end{figure}

{\it Stuart-Landau oscillators.}
The second model we have analysed is a network of Stuart-Landau oscillators
(SL),  
\be   
\dot{w_{i}}=w_{i}-(1+ic_{2})|w_{i}|^{2}w_{i}+g(1+ic_{1}) ( W_i - w_i) \quad ,
\ee
where
$w_i$ is a complex variable, and $W_i = (1/K)\sum_{j=1}^{N}S_{ij}w_{j}$
represents the local field ($S_{ij}$ is defined as
before). Since the local variable is a complex number, 
it is convenient to introduce the global mean field
$W(t) \equiv |{\overline W}|$ (where $|\cdot |$ denotes the modulus operation),
which essentially coincides with the Kuramoto order parameter~\cite{kura}.
The data reported for $\langle W \rangle$ in
Fig.~\ref{sl:1}a reveals the discontinuous emergence of some form
of synchronization (at least for our choice of the parameter
values, $c_{1}=-2$, $c_{2}=3$ and $g=0.47$~\cite{note3}).
More precisely, there exists a finite parameter range
($55<K<85$), where, depending on the initial conditions~\cite{note2},
$\langle W \rangle$ may either vanish or take a finite value.
A more precise characterization of the colletive phase can be
obtained by looking at the variance of the order parameter,
namely $\sigma_{w}^{2}=\langle W^{2} \rangle - \langle W \rangle^{2}$. 
The data in Fig.~\ref{sl:1}b reveals that the discontinuous transition
is accompanied by the birth of temporal fluctuations which increase with the
connectivity. As shown in the inset of Fig.~\ref{sl:1}b, the global attractor
exhibits an irregular dynamics. Moreover, it is remarkable that the attractor is
qualitatively different from the one found in the mean-field version of the
model~\cite{KazumasaPRL2009}, which is centered around $W \simeq 0.35$.

As for LM, we have verified that both below and above the transition region, the
microscopic dynamics is chaotic. In both cases, there is a clear evidence of a convergence 
towards an asymptotic LE spectrum (the LE spectra for $K=10$ are reported in 
Fig.~\ref{sl:1}c). Analogous conclusions can be drawn from the CE spectra 
(see Fig.~\ref{sl:1}d).

\begin{figure}[h]
\begin{center}
\includegraphics*[angle=0,width=8cm]{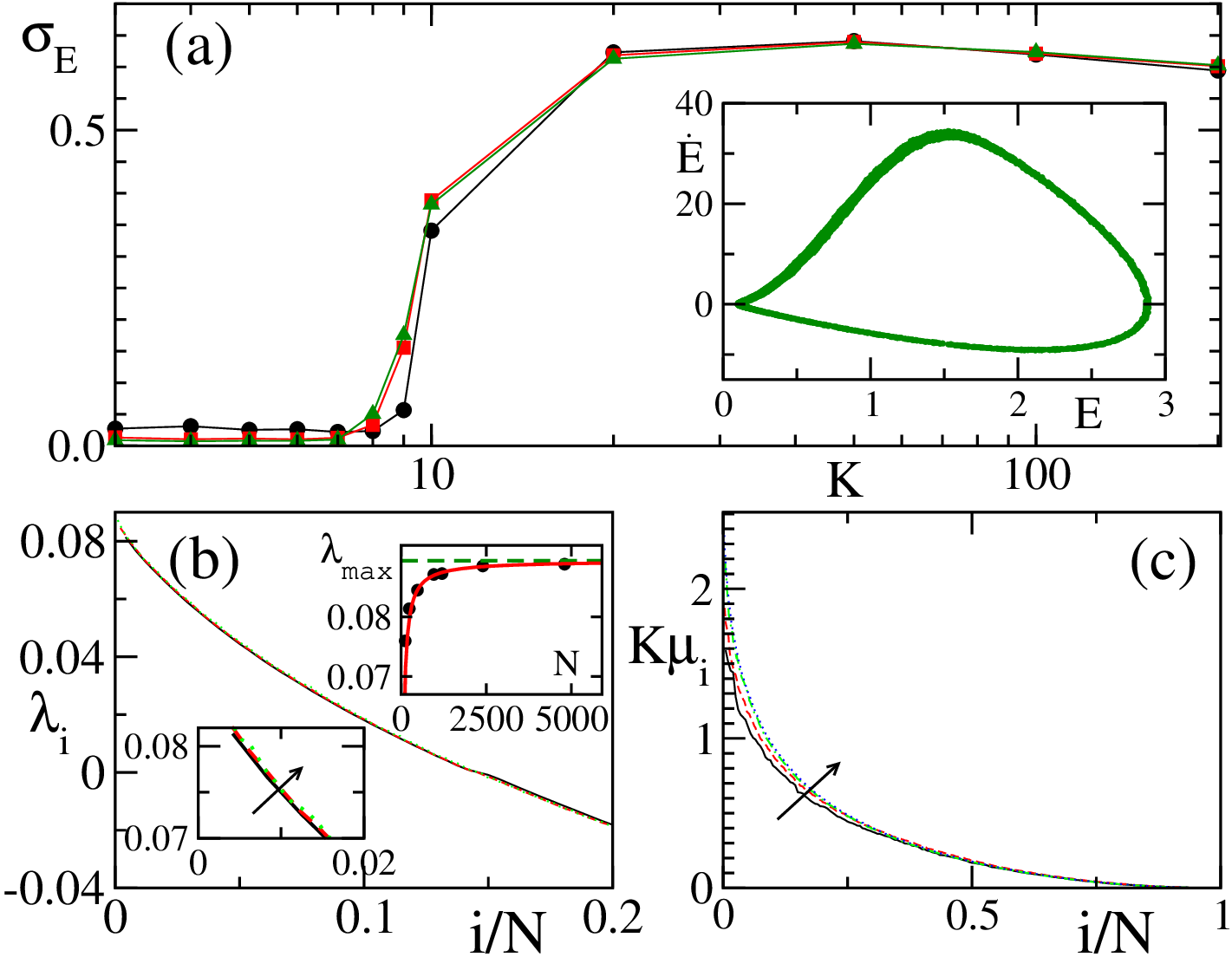}
\end{center}
\caption{(color online) Model LIF.
(a) Standard deviation of the mean field, $\sigma_E$, 
versus $K$ for $N=1,000$ (black) circles, 
$N=5,000$ (red) squares, $N=10,000$ (green) triangles. 
The inset shows the macroscopic attractor
for $N=5,000$ and $K=200$.
(b) LE spectra (in the lower inset a zoom 
of the largest values) for $K=20$ and $N=240-480-960$.
In the upper inset the maximum Lyapunov exponent, $\lambda_{max}$,
versus N is shown, the (red) line represents the nonlinear fit 
$\lambda_{max}=0.0894-2.3562/N$ and the (green) 
dashed line marks the asymptotic value. 
(c) CE spectra (rescaled by $K$) for $K=20$ and $N=200-400-800-1600$.  
} 
\label{lif:1}
\end{figure}

{\it Leaky integrate and fire neurons.} Finally, we have considered LIF
pulse-coupled neurons. They are among the most popular and yet simple models
used in computational neuroscience, the field where understanding the onset of
collective motion is likely to have the deepest impact. 
The evolution equation for the membrane potential $v_i$,
writes as  $\dot{v}_{i}= a -v_{i} + ge_{i}$,
where the local field $e_i$ satisfies the equation~\cite{Abbott_VanVrewswijk1993}
\be
\ddot e_{i} + 2\alpha \dot e_{i} + \alpha^2 e_{i} = \frac{\alpha^{2}}{K} 
\sum_{n|t_n<t} S_{il(n)} \delta(t-t_{n})  \ . 
\ee
Whenever the membrane potential $v_l$ reaches the threshold $v_l=1$, it is
instantaneously reset to the value $v_l=0$ and a so-called  $\alpha$-pulse is
emitted towards the connected  neurons (for more details
see~\cite{Abbott_VanVrewswijk1993}). In this case, we introduce the mean  field
$E= {\overline e}_i$ and the corresponding standard deviation $\sigma_E$. The
model has been simulated for $a=1.3$, excitatory synaptic strength $g=0.2$ and
inverse pulse-width $\alpha=9$. For such values, it is known that in the global
coupling limit, there exists a periodic collective motion accompanied by a
quasi-periodic microscopic dynamics \cite{VanVreeswijk_1996}. For small $K$, the
mean field is constant in the thermodynamic limit, revealing a so-called
asynchronous state, while above a critical value $K_c\approx 9$, it oscillates
periodically, as seen in the inset of Fig.~\ref{lif:1}a. As in the previous
systems, the Lyapunov analysis reveals an extensive behaviour, including the
initial part (see the two insets), where at finite-size corrections of order
$1/N$ are detected. This is to be contrasted with the initial non extensive
``layer" observed in globally coupled systems \cite{KazumasaPRL2011}. Fully
extensive behaviour was already observed for the $\Theta$-neuron model in
\cite{Monteforte_Wolf2010}. Finally, extensivity is confirmed by the orthogonal
decomposition applied to the fluctuations of the local field $e_i$ (see the CE
spectra shown in Fig.~\ref{lif:1}c).

{\it Discussion.} By studying three different models, we find that a
finite connectivity is able to sustain a nontrivial collective motion,
as signalled by the appearance of finite temporal fluctuations of the
mean field. This scenario emerges in LMs, whose entire Lyapunov
spectrum is positive, as well as in LIF neurons that cannot behave
chaotically as stand-alone devices, even when subject to an irregular
forcing. The differences among the various models concern the nature
of the transition (continuous in LMs and LIF neurons, discontinuous and
hysteretic in the SL oscillators) and the complexity of the collective
phase that is periodic in LIF neurons but certainly higher dimensional
in the other two models. It is desirable to trace back analogies and
differences to some general properties of the single systems. In this
perspective, it is natural to assume that a finite connectivity acts as a
mean field accompanied by an effective noise (due to the ``statistical"
differences among the fields seen by the different nodes).
Accordingly, one expects that a small $K$ corresponds to a large noise and is 
thereby unable to maintain a global coherence, as indeed observed. This picture is, 
however, rather qualitative, since the collective motion of SL oscillators
markedly differs from that one generated in the mean-field limit, even when
$K\approx 800$.
Altogether, it is remarkable that a few tens of connections consistently
suffice to sustain a collective motion in such different environments. 

In order to shed further light on the phenomenon, we have added a
spatial structure, by organizing the nodes along a line and
adding finite-distance interactions. When only the latter are present, no
collective motion can arise, because of the low dimensionality of the
lattice~\cite{manneville}. The mixed case corresponds to a ``small-world'' 
arrangement ~\cite{strogatz}; we have imposed that the connectivity 
is the sum of two contributions ($K=2 N_s+k_R$): a number $k_R$ of random 
links and $2 N_S$ links with the nearest nodes. 
The data reported in Fig.~\ref{mappe:1} for $N_s=40$
shows that it is sufficient to add $k_R=20$ links to establish a collective
dynamics; in other words, in the presence of a lattice structure, a lower
number of long-range connections may be necessary, although the overall connectivity
is larger. This observation reveals that the network structure plays a
nontrivial role in determining the number of links that 
can sustain macroscopic motion. 

Next, we comment on the extensive nature of the microscopic motion, a property
that is much less obvious than one could think. In fact, the existence of
a limit Lyapunov spectrum in regular lattices is the natural consequence
of the additivity of the dynamics \cite{grass89}. Imagine to use a hyperplane
$\bf P$ to divide a spatial domain of size $N$ into two subdomains ${\bf S}_1$
and ${\bf S}_2$ of size $N/2$. The mutual interaction between ${\bf S}_1$ and
${\bf S}_2$ is negligible as it affects only the interfacial region around
$\bf P$ (it is a ``surface'' effect). As a result, the entire system can be
approximately seen as the juxtaposition of two independent subsets.
In the case of sparse networks, it is not even obvious how to split them
in two minimally-connected components ${\bf S}_1$, ${\bf S}_2$. This 
problem goes under the name of graph
bipartitioning; it is known to be NP complete, and, more important, the
solution involves a number of connections that is proportional to $N$
itself, whenever $K > 2 \log 2$ \cite{LiaoPRL1987,PercusJMatPh}.
Therefore, the ``interface'' cannot be likened to a ``surface'' and the
evolution is necessarily non-additive. Accordingly, the extensive nature of
the microscopic evolution is due to subtle
properties, that have yet to be clarified.

{\it Conclusions and open problems.} 
In this Letter we have shown that a double-scale (microscopic/macroscopic)
evolution is a generic feature of sparse networks. Altogether, the existence
of a critical connectivity separating asynchronous from coherent 
activity is similar to what experimentally observed in neuronal
cultures~\cite{soriano}. In the perspective of understanding the conditions
for the onset of this behavior, it will be worth exploring the dependence 
on the coupling strength. In particular, in the weak coupling limit,
it might be possible to develop an analytic treatment (as already done
for the synchronization transition in LIF neurons \cite{hansel}),
although some of the temporal complexity might be lost.

\begin{acknowledgments}
AT acknowledges the VELUX Visiting Professor Programme 2011/12 
for the support received during his stay at the University of Aarhus (Denmark).
\end{acknowledgments}


\begin{thebibliography}{xx}
\bibitem{note2} The hysteresis has been detected by simulating a network where
links are first slowly added and then removed.
\end{thebibliography}
\end{document}